# Sequentially Deposited versus Conventional Nonfullerene Organic Solar Cells: Interfacial Trap States, Vertical Stratification, and Exciton Dissociation

*Jiangbin Zhang, Moritz H. Futscher, Vincent Lami, Felix U. Kosasih, Changsoon Cho, Qinying Gu, Aditya Sadhanala, Andrew J. Pearson, Bin Kan, Giorgio Divitini, Xiangjian Wan, Daniel Credgington, Neil C. Greenham, Yongsheng Chen, Caterina Ducati, Bruno Ehrler, Yana Vaynzof\*, Richard H. Friend, and Artem A. Bakulin\**

J. Zhang, Dr. C. Cho, Q. Gu, Dr. A. Sadhanala, Dr. A. J. Pearson, Dr. D. Credgington, Prof. N. C. Greenham, Prof. R. H. Friend;
Cavendish Laboratory, University of Cambridge, JJ Thomson Avenue, Cambridge CB3 0HE, United Kingdom;
J. Zhang, Dr. A. A. Bakulin;
Department of Chemistry, Imperial College London, London SW7 2AZ, United Kingdom;
Email: a.bakulin@imperial.ac.uk
M. H. Futscher, Dr. B. Ehrler;
Center for Nanophotonics, AMOLF, Science Park 104, 1098 XG Amsterdam, The Netherlands;
Email: vaynzof@uni-heidelberg.de
V. Lami, Prof. Y. Vaynzof;
Kirchhoff Institute for Physics and the Centre for Advanced Materials, Heidelberg University, Im Neuenheimer Feld 227, 69120 Heidelberg, Germany;
Prof. Y. Vaynzof;
Dresden Integrated Centre for Applied Physics and Photonic Materials (IAPP) and Centre for Advancing Electronics Dresden (CFAED), Technical University of Dresden, Nöthnitzer Straße 61, 01187 Dresden;
F. U. Kosasih, Dr. G. Divitini, Prof. C. Ducati;
Department of Materials Science & Metallurgy, University of Cambridge, 27 Charles Babbage Road, Cambridge CB3 0FS, United Kingdom;
Dr. C. Cho;
School of Electrical Engineering, Korea Advanced Institute of Science and Technology (KAIST), Daejeon 34141, Republic of Korea;
Dr. B. Kan, Prof. X. Wan, Prof. Y. Chen;
The Centre of Nanoscale Science and Technology and Key Laboratory of Functional Polymer Materials, State Key Laboratory and Institute of Elemento-Organic Chemistry, College of Chemistry, Nankai University, Tianjin, 300071, China;

*Abstract*: Bulk-heterojunction (BHJ) non-fullerene organic solar cells prepared from sequentially deposited donor and acceptor layers (sq-BHJ) have recently been promising to be highly efficient, environmentally friendly, and compatible with large area and roll-to-roll fabrication. However, the related photophysics at donor-acceptor interface and the vertical heterogeneity of donor-acceptor distribution, critical for exciton dissociation and device performance, are largely unexplored. Herein, steady-state and time-resolved optical and electrical techniques are employed to characterize the interfacial trap states. Correlating with the luminescent efficiency of interfacial states and its non-radiative recombination, interfacial trap states are characterized to be about 50% more populated in the sq-BHJ devices than the as-cast BHJ (c-BHJ), which probably limits the device voltage output. Cross-sectional energy-dispersive X-ray spectroscopy and ultraviolet photoemission spectroscopy depth profiling directly visualize the donor-acceptor vertical stratification with a precision of 1-2 nm. From the proposed "needle" model, the high exciton dissociation efficiency is rationalized. Our study highlights the promise of sequential deposition to fabricate efficient solar cells, and points towards improving the voltage output and overall device performance via eliminating interfacial trap states.

**1. Introduction**

Organic solar cells (OSCs), made from solution-processable carbon-based materials, have the potential to be flexible, light-weight and low-cost.[1] Using fullerene and its derivatives as benchmark electron-accepting materials, tremendous efforts in developing electron-donating polymers and small molecules, particularly low-bandgap materials, have taken the device power conversion efficiency (PCE) over 10%.[2–9] The drawbacks of fullerene, such as being expensive, unstable and not absorptive in the near-IR region have largely been overcome by the fast development of small-molecule acceptors, the so-called non-fullerene acceptors (NFAs).[10–24] These molecules exhibit tunable absorption and energy levels, and contribute to efficient photocurrent generation even at a negligible driving force.[25–27] As such, PCEs of binary and tandem devices have reached over 16% and 17.3%, respectively.[28,29]

The efficiency of the planar heterojunction (PHJ) devices, when donor and acceptor layers are placed on top of each other, is mainly limited by the so-called "exciton bottleneck", the competition requirement for efficient optical absorption and limited exciton diffusion.[30] A major breakthrough was the invention of bulk-heterojunction (BHJ) – an inter-penetrating donor and acceptor network.[31,32] This structure can be easily obtained by spin-coating the blended donor and acceptor solutions, but the morphology is very sensitive to the materials and processing conditions, such as the blend ratio, solvent and solvent additives as well as the thermal and solvent annealing processes.[33] An intermediate active layer nanomorphology between PHJ and BHJ is termed the graded bulk heterojunction (GBHJ).[34] The gradient morphology contributes to increased exciton dissociation efficiency relative to the PHJ and an enhanced charge collection efficiency compared to a uniformly mixed BHJ.[34,35] The morphology of the GBHJ can be controlled in vacuum deposited binary films where the ratio of donor/acceptor deposition rate is ramped linearly, or a stack of thin layers with varied donor-acceptor concentration ratios.[35] Experimental methods to prepare the GBHJ via solution processing are less straightforward, and can involve manipulating the surface energy of substrates, substrate temperature, solvent fluxing, and graded nanoparticle layers.[36–39]

A method to prepare GBHJ originating from fullerene-based cells, called sequential deposition (sq-BHJ), or layer-by-layer approach attracted much attention last year in developing high-efficiency NFA-based OSCs.[40–47] To better control the phase separation, Hou *et al.* used a mixed solvent for a new polymer in combination with a high-performance NFA where the interdiffusion was controlled by the amount of a second solvent. This exercise led to an efficiency at 13% for sq-BHJ devices, higher than 11.8% obtained by the one-step processing.[48] Huang *et al.* and Min *et al* successfully applied this method to fabricate large-area (1 cm$^2$) devices with a performance of over 10% and improved device stability.[49,50] Yang *et al.* fabricated ternary blends in which a BHJ was mixed with a new donor or acceptor layer.[51] In the same period, our group found that sequentially depositing the donor and acceptor layers led to a high efficiency (>10%), comparable to the as-cast one-step formation of BHJ (c-BHJ) using novel NFAs.[52] Such advancements in device efficiency, stability, green-solvent and large-area processing make this sequential deposition method universal and attractive.

So far, most studies on sq-BHJ systems have focused on device performance rather than a detailed mechanistic study of the underlying photophysics. The reasons and mechanism for the comparable performance need to be understood, and obvious questions remain behind sq-BHJ functionality. For example, to realize high (close to unity) charge generation efficiency in sq-BHJ devices, most excitons must be separated at the donor-acceptor (D-A) interface. Characterizing this process is a prior to understanding efficient device operation. In this work, we focus on interfacial properties in sq-BHJ together with morphological characterizations to study their relationship with the initial exciton dissociation and device performance. Using a range of spectroscopic techniques, we focus on the interfacial states at the D-A interfaces in blends prepared by sequential deposition as well as as-cast one-step methods, and correlate our observations with the device performance. To directly visualize the vertical stratification, we characterize the D-A vertical distribution using cross-sectional transmission electron microscopy-energy dispersive X-ray spectroscopy (TEM-EDX) and ultraviolet photoemission spectroscopy (UPS) depth profiling. To understand the effect of D-A distribution on exciton dissociation, a "needle" model is proposed to simulate the structure of sq-BHJ compared with a "cubic" structure for c-BHJ.

## 2. Results and Discussion

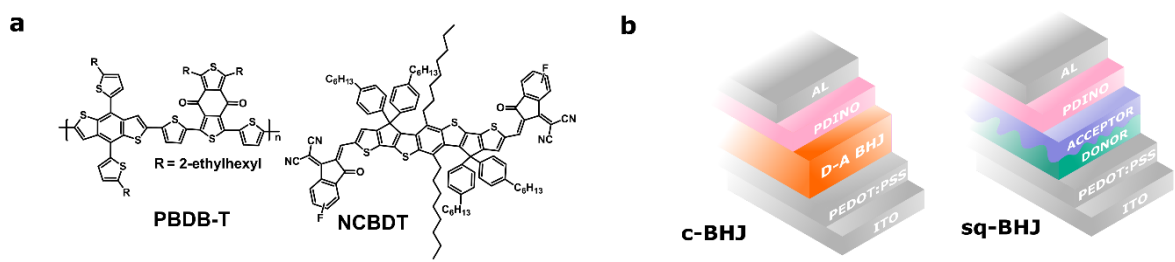

**Figure 1.** (a) Chemical structures of PBDB-T (donor) and NCBDT (acceptor). (b) Device configuration in which the photoactive layer is based on a c-BHJ (one-step processing) or sq-BHJ architecture (sequential deposition). PEDOT:PSS and PDINO are used as the hole transport layer and the electron transport layer, respectively.

**Figure 1**(a) shows the molecular structures of donor (poly[(2,6-(4,8-bis(5-(2-ethylhexyl)thiophen-2-yl)-benzo[1,2-b:4,5-b']dithiophene))-alt-(5,5-(1',3'-di-2-thienyl-5',7'-bis(2-ethylhexyl)benzo[1',2'-c:4',5'-c']dithiophene-4,8-dione)], PBDB-T) and acceptor (NCBDT) molecules. PBDB-T, first synthesized by the Hou group, is a benchmark polymer for NFA-based blends.[53] NCBDT is a benzodithiophene-core based small molecule. We previously demonstrated high performance devices (PCE > 12%) using this D-A combination.[21]

Most recently, we used this compound as active materials to fabricate sq-BHJ devices and achieved efficiencies in excess of 10%.[52] We also observed that the open-circuit voltage ($V_{OC}$) of sq-BHJ devices was ~30 meV smaller than that of the c-BHJ devices. Such loss is correlated with increased non-radiative recombination, observed from their electroluminescence (EL) efficiency.[52] In comparing devices of the same material combination, the most probable explanation for the difference in EL efficiency is the relative population of trap states. Because the donor and acceptor stoichiometry is similar in both blends, we expect that the trap states are interfacial and depends on the processing history.[54]

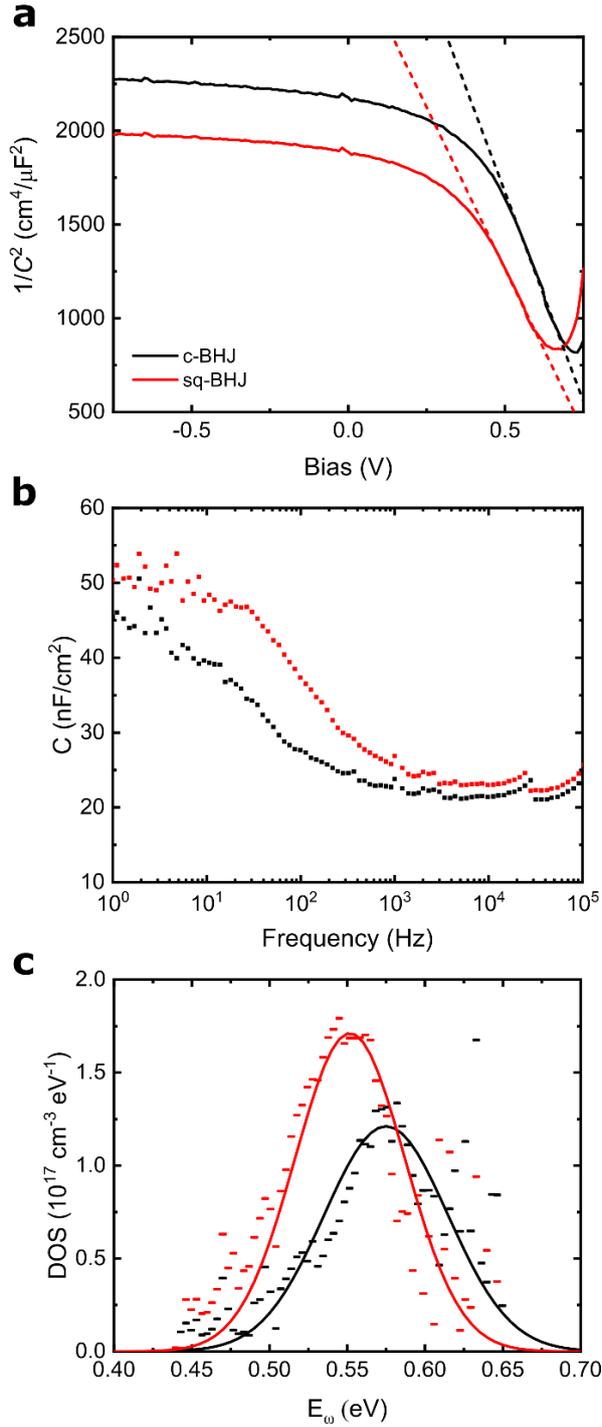

**Figure 2.** (a) Mott-Schottky plot measured at 10 kHz. The linear fit reveals the doping density and the built-in potential. (b) Capacitance spectra measured at zero bias. (c) Density of trap states (DOS) calculated using the capacitance spectra shown in (b). The continuous line corresponds to the fit of a Gaussian defect distribution.

We characterize these interfacial trap states using capacitance measurements. **Figure 2** (a) shows the capacitance as a function of voltage, where a plateau at low voltages indicates full depletion at the short-circuit condition. At low forward bias, we observe a capacitance increase which is correlated to a decrease in depletion-layer width. This change in capacitance can be approximated by the Mott-Schottky relation as

$$C^{-2} = \frac{2}{\varepsilon_0 \varepsilon \, qNA^2} (V_{bi} - V) \tag{1}$$

where $\varepsilon_0$ is the vacuum permittivity, $\varepsilon$ the blend permittivity, $q$ the elementary charge, $N$ the doping density, $A$ the device active area, $V_{bi}$ the built-in potential, and $V$ the external bias. From the Mott-Schottky plot we obtained a doping density of $(7.3 \pm 0.6) \times 10^{16}$ cm$^{-3}$ for the sq-BHJ, slightly higher than the value of $(6.2 \pm 0.5) \times 10^{16}$ cm$^{-3}$ for the c-BHJ, and a built-in potential of $0.87 \pm 0.02$ V and a blend permittivity of $2.4 \pm 0.2$ V for both the c-BHJ and the sq-BHJ. To quantify the density and energetics of trap states in both devices, we measured the capacitance as a function of frequency at zero bias in the dark (see Figure 2 (b)). At low frequencies, we observe an increase in capacitance due to charging and discharging of defect states.[55] At high frequencies, the defects cannot follow the applied AC signal. Using these capacitance spectra, the defect distribution can be estimated as

$$N_T(E_\omega) = -\frac{V_{bi}\omega}{qwk_BT} \frac{dC}{d\omega} \tag{2}$$

where $E_\omega$ is the demarcation energy, $k_B$ the Boltzmann constant, $w$ the depletion width, and $\omega$ the modulation frequency.[55] $E_\omega$ is calculated as

$$E_\omega = k_BT \ln\left(\frac{\omega_0}{\omega}\right) \tag{3}$$

where $\omega_0$ is the attempt-to-escape frequency. Assuming a typical attempt-to-escape frequency of $10^{12}$ s$^{-1}$, we find a Gaussian density of trap states centred between 0.5 eV and 0.6 eV of $(1.7 \pm 0.1) \times 10^{17}$ cm$^{-3}$ for the sq-BHJ, around 50% higher than the value of $(1.2 \pm 0.1) \times 10^{17}$ cm$^{-3}$ for the c-BHJ (see Figure 2 (c)).[56] The fitting is a weighted fit that gives less weight to the low frequency region due to a large error in capacitance. The density of the trap states is higher than the acceptor density obtained from the Mott-Schottky analysis because the capacitance-voltage characteristics were measured at a frequency (10 kHz) at which the defects cannot follow the applied AC signal. We furthermore find that density of trap states of the c-BHJ (40 meV) is broader than that of the sq-BHJ (35 meV), suggesting less molecular disorder at the interface in the latter case.

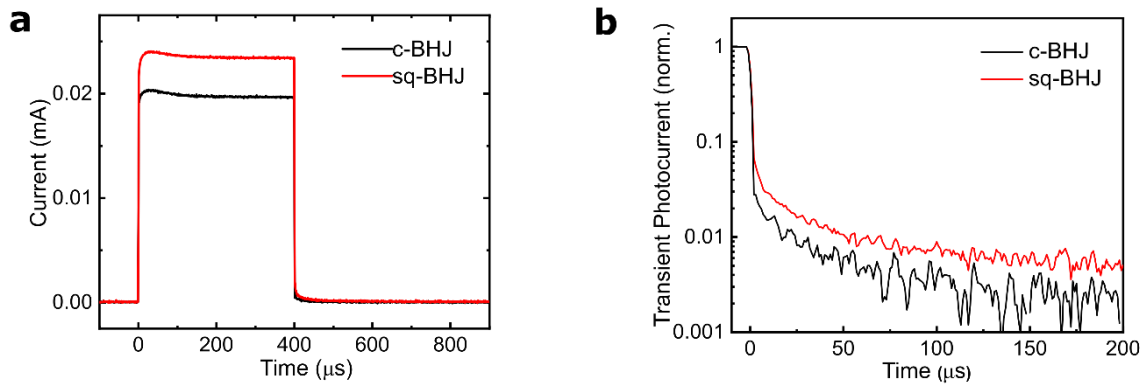

**Figure 3.** (a) Decay curves from transient photocurrent spectroscopy with excitation at 460 nm. (b) Normalized photocurrent decay in the log scale with the time zero shifted by around 400 µs relative to (a).

We also characterize the trap states from transient photocurrent measurements. **Figure 3** (a) shows a larger steady-state current between 200 and 400 µs for the sq-BHJ OSC. The higher photocurrent is due to its higher EQE of the sq-BHJ device at the illumination wavelength (460 nm).[52] The signal is flatted after 200 µs, which reflects an equilibrium between the trapping and detrapping of free carriers. When the light is switched off (400 µs after the initial excitation), the trapping channel is stopped and only the detrapping of free carriers contributes to the decay curve.[57] The initial fast decay in the curve is due to charge collection. Following the normalization, the relative amplitude or the area below the decay curve is independent of the carrier density, and thus represents the relative density of trap states. From Figure 3(b), we observe a larger area below the decay curve for the sq-BHJ device, which we assign to a higher relative density of trap sites, agreeing with the capacitance measurements. We note that further evidence may come from dark current-density-voltage measurements (**Figure S1**), which also supports that more trap states are present in the sq-BHJ devices.

Trap states have been shown to act as non-radiative recombination centers in Shockley-Read-Hall type recombination, resulting in a lowering of the quasi-Fermi level of the electrons and a limitation in the $V_{OC}$.[58,59] The higher density of trap states in sq-BHJ devices are probably related to their lower $EQE_{EL}$ and slightly lower $V_{OC}$ compared with c-BHJ devices.[52] We note that sequential deposition generally leads to more stable devices.[40,49] However in our blends, under light illumination, generation of deep trap states may exacerbate the degradation of solar cells. Such trap states might originate from a fast solvent evaporation during spin-coating and are probably morphology dependent. Future effort is needed to eliminate these trap states to improve the performance of devices prepared by the sequential deposition method.

From the previous neutron scattering modelling, the morphology of the sq-BHJ is more akin to a homogeneous BHJ than a two-layer PHJ.[52] The detailed morphology distribution of sq-BHJ probably lies between BHJ and PHJ, but its 3D morphology at nm length scale is difficult to determine from experiments. Initial evidence for D:A intermixing throughout the sq-BHJ blend was obtained from XPS measurements. As NCBDT contains N and F atoms and PBDB-T does not, these elements can be used as chemical markers for the presence of the acceptor molecules. **Figure 4 (a)** presents XPS spectra measured on c-BHJ and sq-BHJ films, providing an insight into the elements present at the top ~10 nm of each film surface. The spectrum for the sq-BHJ sample contains more intense F 1s and N 1s features (See **Table 1** for atom% values), indicating surface enrichment of the acceptor at the surface.

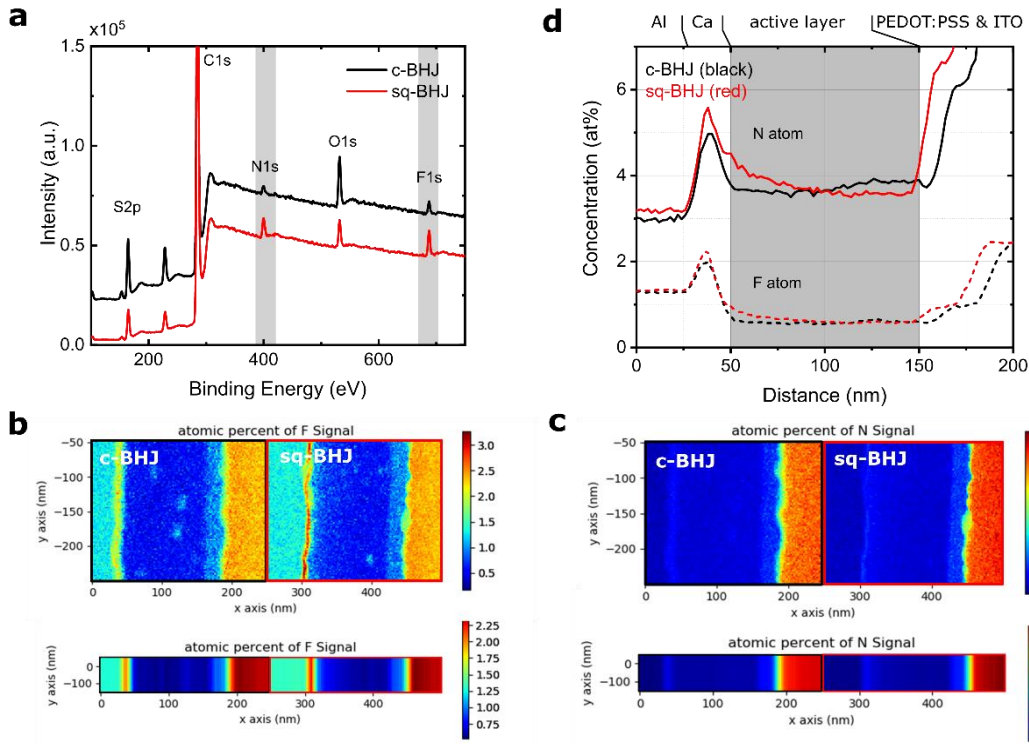

**Figure 4.** (a) XPS on ITO/PEDOT:PSS/active layer. (b-c) Maps of N and F concentrations as measured with STEM-EDX. (d) The N and F concentration plotted from the Al layer (x = 0 nm) up to ITO (x = 200 nm). This information is averaged from vertical slices in (b-c); the grey area is determined to be the active layer. The peaks around x = 40 nm were caused by partial overlap with the very intense O $K_\alpha$ peak at the Ca layer there. In EDX measurements, PDINO is replaced with 10-nm Ca to reduce the possible influence of PDINO on nitrogen concentration.

**Table 1.** Atomic concentration of the top surface of c-BHJ and sq-BHJ devices measured by XPS. The device structure is ITO/PEDOT:PSS/active layer.

| Name | Position (eV) | c-BHJ atomic concentration (%) | Sq-BHJ atomic concentration (%) |
|---|---|---|---|
| O 1s | 532.22 | 4.12 | 2.12 |
| C 1s | 285.22 | 86.74 | 90 |
| **N 1s** | **399.22** | **0.73** | **2.56** |
| **F 1s** | **687.22** | **0.53** | **1.22** |
| S 2p | 165.22 | 7.88 | 4.10 |

To confirm our XPS results, we prepared cross-sectional lamellae (~70 nm) of full devices using a focused ion beam miller for transmission electron microscopy (TEM) characterization. Unfortunately, we were not able to observe noticeable differences in D:A distribution from bright-field as well as high resolution TEM imaging (**Figure S2**). We then switched our TEM to scanning mode and used energy dispersive X-ray spectroscopy (STEM-EDX) to map the elemental distribution in the c-BHJ and sq-BHJ blends. For the EDX measurement, the device configuration is ITO/PEDOT:PSS/active layer/Ca (10 nm)/Al, where PDINO with nitrogen inside was replaced with the 10-nm Ca layer. After acquiring EDX spectrum images of c-BHJ and sq-BHJ layers, we performed principal component analysis (PCA) in HyperSpy[60] to denoise the dataset from which we finally produced semi-quantitative elemental maps. While EDX is generally not capable of measuring concentrations of light elements to a very high accuracy, the elemental distribution trends shown in the maps are clear.[61] Considering the 2D map showing the elemental distributions of F and N in Figure 4 (b-c), we averaged the signal vertically, and obtained a 1D line in Figure 4 (d). We note that the small peak in N and F concentration around x = 40 nm, supposedly for Ca layer, is possibly due to an intense signal for O-K$\alpha$ at 525 eV which also spreads into N-K$\alpha$ (392 eV) and F-K$\alpha$ (677 eV) after decomposition. The strong rise of nitrogen signal after 150 nm might be caused by some impurities in the ITO layer. We determined the active layer region as highlighted in grey in Figure 4 (d). The thicknesses of c-BHJ and sq-BHJ active layer agrees with previous thickness measurements using atomic-force microscope.[52] For the c-BHJ layer, the F distribution is not uniform, but rather shows a gradual increase in the bottom half of the active layer within 10% variation and the N distribution changes following a similar pattern. This variation is probably caused by two acceptor clusters seen in Figure 4 (b), and the concentration at both ends is quite similar, in agreement with the XPS results. We thus concluded that c-BHJ film is fairly homogenous. In contrast, in sq-BHJ films shown in Figure 4 (c), both N and F distribution shows a gradual decrease in concentration (~15% for N and ~30% for F) from the top surface of the active layer up to about half of the layer's thickness. These results support the conclusion that the top half of the film exhibits a gradual change in the D-A composition, while the bottom half of the film has a homogenous D and A distribution.

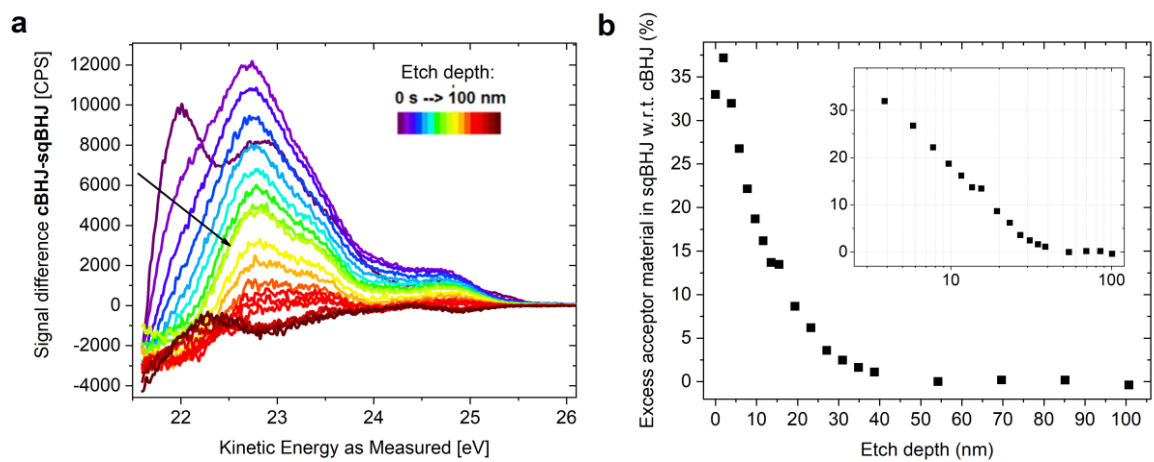

**Figure 5.** (a) The kinetics energy spectra at different etching depths, measured using depth-profile UPS. The resolution is 1-2 nm. (b) The excess acceptor material in sq-BHJ compared to c-BHJ with a function of the etching depth. The data is achived assuming a uniform mixing in the c-BHJ at a ratio of 5:4.

To visualize the vertical distribution further, we employed a newly developed technique: UPS depth profiling.[62] UPS measures the kinetic energy spectrum of emitted photoelectrons after absorbing ultraviolet photons, and thus determines the occupied molecular orbital energies, and the density of states (DOS) in the valence band region. Depth profiling uses Argon (Ar) ion cluster sputtering which does not induce damage to the electronic and chemical structures of the organic materials.[63] The combination of Ar cluster etching with the highly surface-sensitive UPS offers a superior vertical resolution of 1-2 nm, surpassing the capabilities of traditional XPS depth profiling (normally 5-10 nm). We first probe the spectra of pure PBDB-T and NCBDT films (**Figure S3**), which show different distribution of filled states and will be later be used for fitting. The results of UPS depth profiling for both c-BHJ and sq-BHJ are shown in **Figure S4**. On the one hand, the change of the DOS over the entire c-BHJ active layer is not significant, except for a small variation at the very surface, probably induced by surface contamination and by a slightly shifted energetics. On the other hand, the DOS of the sq-BHJ sample shows a continuous change in the top half of the film. Four representative depths are also shown in **Figure S5**, showing that the difference between spectral slices from the c-BHJ and the sq-BHJ vanishes with the increasing etch depth. Since we are primarily interested in these differences between the c-BHJ and sq-BHJ films, we calculated their UPS signal difference for each measured depth as shown in **Figure 5 (a)**. The spectral shape of this difference spectrum (c-BHJ minus sq-BHJ) remains almost the same for each depth, except for the very first spectrum showing small variations at a kinetic energy of ~22 eV due to surface effects. As the spectral shape stays the same, the magnitude of the spectrum (taken as the integral of the signal) is a measure for the difference in the amount of A (or D) in the c-BHJ and in the sq-BHJ, as described in detail in **Supplementary Note 1**. We thus can determine the relative excess of A in the sq-BHJ versus the c-BHJ. However, the absolute values still need a calibration. To quantify this excess, we fitted the very top surface of both c-BHJ and sq-BHJ using the spectra obtained from pure PBDB-T and NCBDT films (Figure S3) with the fits shown in **Figure S6**. From this direct comparison, the excess of acceptor material on the very top surface is found to be ~33%, assuming a homogenous D:A ratio of 5:4 in the c-BHJ film. This is summarized in Figure 5 (b), confirming that the upper half (top ~40 nm) of the sq-BHJ film exhibits a gradually decreasing amount of excess acceptor material, while the bottom half is compositionally equivalent to the c-BHJ. This vertical trend is in excellent agreement with the results of the EDX measurements shown above.

The vertical stratification of binary composition greatly influences charge recombination and transport and is of great relevance to device performance.[64] We note that the view on the resultant vertical phase separation in sq-BHJ films, whether homogeneous or inhomogeneous, is actually not convergent,[65–68] largely depending on the characterisation methods. Using cross-sectional EDX and UPS depth profiling, we directly visualize the vertical phase separation of donor and acceptor and our result is important for studying the morphology in sq-BHJ devices.

Using the sequential deposition method, the acceptor material is enriched on the top layer. This structure is beneficial for the charge transport in the regular device architecture, but actually not good for the inverted devices. To show this effect, we fabricated inverted devices using c-BHJ and sq-BHJ active layers. Their photovoltaic performance is summarized in **Table 2**. We find that sq-BHJ is more sensitive to the device structure, as the device average PCE dropped by ~20% in contrast to ~4% for the c-BHJ layer. To take advantage of the field distribution, we find that the larger bandgap material in D:A blends is preferred to be positioned near the metal electrode to increase light absorption. We note that recent sq-BHJ blends are disadvantageous in this aspect, and we believe such strategy may further improve the device performance of sq-BHJ devices. A more detailed discussion is given in the supporting information.

**Table 2. Photovoltaic performance of regular and inverted PBDB-T:NCBDT devices prepared with one-step formation of BHJ and sequential deposition without post-annealing or solvent additives.**

| Device structure | Active layer layout | $V_{OC}$ [V] | $J_{SC}$ [mA cm$^{-2}$] | FF [%] | PCE [%] |
|---|---|---|---|---|---|
| **Regular** | c-BHJ[a] | 0.847 (0.842±0.003) | 18.64 (18.32±0.20) | 64.6 (63.5±0.5) | 10.19 (10.05±0.12) |
| | sq-BHJ[a] | 0.824 (0.820±0.003) | 19.45 (19.14±0.15) | 62.9 (61.8±0.6) | 10.04 (9.70±0.24) |
| **Inverted** | c-BHJ | 0.855 (0.847±0.006) | 20.13 (18.56±0.88) | 61.65 (61.21±1.78) | 10.62 (9.63±0.54) |
| | Sq-BHJ | 0.839 (0.814±0.010) | 18.39 (17.24±1.22) | 61.50 (54.56±2.59) | 9.49 (7.67±0.84) |

[a] data from reference.[52]

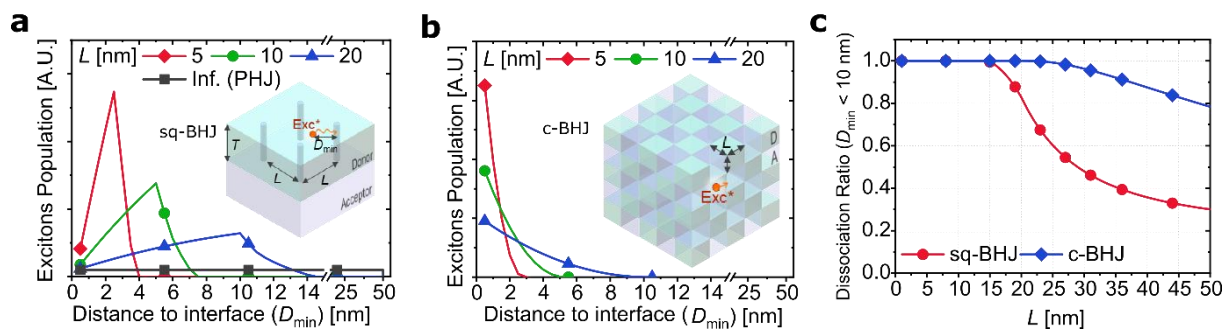

**Figure 6**. A simple model of D:A morphology to simulate the exciton dissociation in sq-BHJ, c-BHJ and PHJ. (a, b) Modelled exciton population ratio as a function of the minimum distance to donor/acceptor interface (*D*) in the (a) sq-BHJ and (b) c-BHJ. For the sq-BHJ, donor and acceptor are assumed to be thin-films sequentially deposited with a thickness of *T* = 50 nm. Acceptor is assumed to be mixed to donor in a shape of ultrathin columns, having a period of *L* and height equal to *T*. For the c-BHJ, donor and acceptor are equally mixed as shown in the inset, where each cube has an edge length of *L*. It should be noted that the sq-BHJ with infinite *L* corresponds to the PHJ. Excitons are assumed to be uniformly generated in the donor region and those in the acceptor region can be calculated in the same way. (c) Exciton dissociation ratio of the sq-BHJ and c-BHJ structures as a function of $L$, assuming that excitons with $D_{min}$ < 10 nm are fully dissociated.

For the exciton dissociation step, here we propose a simple model for sq-BHJ structure assuming that acceptor phase columns with ignorable volume like a "needle" are grown in the planar donor layer. We note that the needle model is an approximation and the actual morphology might look more similar to a homogenous mixture with a certain gradient ratio along the vertical direction. Both needle and homogeneous-mixing models would be in agreement with STEM-EDX and depth-profile XPS data, because both techniques measure the overage over the beam area (~mm$^2$) which exceeds the scale of phase separation (~10 nm). Same time the needle model we propose is probably a better approximation to describe the following aspects: (i) the likely partial de-mixing of donor and acceptor leading to relatively pure phases; (ii) the typical spatial scale at which excitons diffuse before they dissociate. Inset in **Figure 6** (a) shows its schematic where the distance for an exciton to meet the interface ($D_{min}$) is the minimum horizontal distance to the column or the vertical distance to the planar interface. From the calculated probability distribution, the most probable $D_{min}$ increases with the distance between the columns ($L$). In the extreme case when $L$ is infinite, this structure turns into a PHJ. As excitons in the PHJ can only dissociate through the D-A interface, the population function is uniform over $D_{min} = 0 \sim T$ ($T = 50$ nm), while excitons in the sq-BHJ can be more easily dissociated through the columns. Figure 6 (b) shows the "cubic" model for the c-BHJ structure, where excitons are generated in small cubes with a size of $L$, having a large area of interfaces per volume. The $D_{min}$ here is defined as the minimal distance to the surrounding surfaces of another material. Considering the realistic exciton diffusion length of organic materials, we assume that excitons within $D_{min}$ < 10 nm are fully dissociated. By comparing the fraction of dissociated excitons to those with $D_{min}$ > 10, we calculate the exciton dissociation efficiency shown in Figure 6 (c). C-BHJ (blue) achieves the unity dissociation efficiency for $L$ < 20 nm, and the efficiency is maintained to be >78% with a very large $L$ of 50 nm, showing efficient exciton dissociation with a relatively small dependence on the morphology. On the other hand, the optimization of morphology is shown to be more important in the sq-BHJ, as depicted by the rapid drop of dissociation efficiency for an increased $L$. Such sharp dependency on morphology can be attributed to the relatively small interfacial area of the structure, where the excitons have no alternative path to be dissociated if a column moves far apart. With a well-controlled morphology, the exciton dissociation efficiency of the sq-BHJ can also reach 100%, when $L < 14.2$ nm, while such efficiency of PHJ is only 20%. To secure efficient exciton dissociation, we can infer that $L$ should be small and this agrees with the high device performance already achieved in sq-BHJ devices. Such morphology with a small $L$ is highly possible in sq-BHJ, considering its rough surface and the possibility of vertical heterogeneity. Thus, our model provides a simple picture to justify the promise of efficient exciton dissociation in sq-BHJ. We note that this model can be easily modified when knowing the phase separation in more detail.

## 3. Conclusion

In summary, we have characterized the interfacial trap states at donor-acceptor interface and the vertical heterogeneity of donor-acceptor distribution in a highly efficient polymer:NFA blend prepared by sequential deposition and conventional one-step processing methods. With capacitance measurements and transient photocurrent spectroscopy, we find around 50% more trap states in sq-BHJ devices compared to c-BHJ ones. These trap states at the interface can adversely influence the device performance, such as the non-radiative recombination, limiting voltage improvement. The vertical stratification is directly visualized using two advanced techniques, cross-sectional TEM-EDX and depth-profile UPS, supporting gradual D-A composition change in top half of the film and a uniform distribution in the bottom half. Our proposed simple model to simulate the sq-BHJ structure demonstrates that sq-BHJ devices can achieve unity exciton dissociation without such strong morphological requirements as in traditional BHJ systems. Our results highlight the need to eliminate these trap states to achieve higher $V_{OC}$ and PCE in sq-BHJ devices.

## 4. Experimental Section

*Materials.* PBDB-T was purchased from Ossila (M1002). NCBDT was synthesized using the procedure reported elsewhere.[21] Chloroform, dichloromethane (DCM) and zinc oxide (ZnO) were bought from Sigma Aldrich.

*OPV device fabrication.* The device structure was glass/ITO/PEDOT:PSS (poly(3,4-ethylenedioxythiophene) polystyrene sulfonate)/active layer/PDINO (perylene diimide functionalized with amino N-oxide)/Al. The glass substrate with ITO was cleaned sequentially by deionized water, acetone and isopropyl alcohol under ultrasonication for 10 min each. The subsequent PEDOT:PSS layer was spin-coated at 5000 RPM for 45 s, and then baked at 150 ºC for 20 min in ambient atmosphere. For the sq-BHJ film, the donor layer was deposited from 6 mg/ml solution in chloroform at 1900 RPM for 20 s, and the subsequent acceptor layer was cast from DCM solution (~60 uL, 6 mg mL$^{-1}$) at 2500 RPM for 40 s right before spin-coating. PDINO (1 mg mL$^{-1}$ in $CH_3OH$) was spin-coated on the active layer at 3000 RPM for 40 s. Finally, a 100 nm Al layer was deposited under high vacuum. The effective area of each cell was 4.5 mm$^2$. For the inverted devices, the device structure was glass/ITO/ZnO/active layer/$MoO_3$/Ag. The ZnO precursor was prepared from dissolving zinc acetate dehydrate (1.098 g) in 2-methoxyethanol (10 mL) mixed with ethanolamine (301.8 µL) as a stabilizer. The precursor was stirred on a hot plate at 1000 RPM at 60 ºC for at least 2 hours. The fully dissolved solution was filtered using 0.2 µm PTFE filters, and then the ZnO layer was spin-coated at 3000 RPM for 60 s before being baked at 80 ºC for 10 min and 130 ºC for 1 hour in ambient atmosphere, resulting in a ~30 nm-thick film. The active layer was deposited in the same way as the conventional devices and a 10 nm $MoO_3$ as well as a 100 nm silver layer was deposited under high vacuum.

*Transient photocurrent spectroscopy.* A 465 nm light-emitting diode (LED465E, Thorlabs) was used as the light source for transient experiments, connected to an Agilent 33500B wavefunction generator and a purpose-built low-noise power supply. Solar cell transients were recorded by connecting the device to a Tektronix DPO 3032 oscilloscope. For TPC measurements, the device was connected to the 50 Ω input of the oscilloscope via a custom trans-impedance amplifier. A custom-written LabVIEW VI was used for instrument control and data acquisition.

*Capacitance measurements.* Capacitance measurements were performed at a pressure below $3\times10^{-6}$ mbar in the dark at 300 K with an AC perturbation of 20 mV. For the fitting, the thickness of the active layer thickness was set to $(100 \pm 5)$ nm for the c-BHJ and $(90 \pm 5)$ nm for the sq-BHJ devices.

*X-ray photoelectron spectroscopy.* The structure of samples for XPS measurements was ITO/PEDOT:PSS/active layer, using the same procedures as for device fabrication. Some films were ready for further measurements, while other films were cut into ~3 mm × 3 mm squares and immersed in water. The floating pieces were then carefully transferred to silicon wafer substrates with and without flipping. The samples (either glass substrates or silicon wafer) were then transferred to an ultrahigh vacuum (UHV) chamber (ESCALAB 250Xi) for XPS measurements, using an XR6 monochromated Al K Alpha X-ray source ($hv$ = 1486.68 eV) with a 400 µm spot size and 200 eV pass energy.

*Depth-profile ultraviolet photoelectron spectroscopy.* The samples were prepared in the same way as the corresponding c-BHJ and sq-BHJ photovoltaic devices. After preparation, the samples were stored in $N_2$ and afterwards transferred into an ultrahigh vacuum (UHV) chamber of a photoelectron spectroscopy (PES) system (Thermo Scientific ESCALAB 250Xi) for measurements. The samples were exposed to air only for a short time span of approximately 30 seconds during this transfer. All measurements were performed in the dark. UPS measurements were carried out using a double-differentially pumped He discharge lamp ($hv$ = 21.22 eV) with a pass energy of 2 eV and a bias at -5 V. Etching was performed using an Argon cluster (MAGCIS) source with a cluster energy of 4000 eV and a raster size of 2.5×2.5 mm$^2$. Otherwise, UPS depth profiling was performed the same way described in Lami et. al.[62]

*Scanning transmission electron microscopy – energy dispersive X-ray spectroscopy.* The device layout is ITO/PEDOT:PSS/active layer/Ca(10 nm)/Al (100 nm). Sample lamellae were prepared with a FEI Helios Nanolab Dualbeam focused ion beam/scanning electron microscope. STEM imaging and STEM-EDX were conducted in a FEI Tecnai Osiris TEM fitted with a Schottky X-FEG gun and operated at 80 kV acceleration voltage. EDX spectrum images were acquired using a Bruker Super-X detector with a collection solid angle of ~0.9 sr and spatial resolution of 2 nm/pixel. EDX data were obtained with Tecnai Imaging and Analysis software and analysed in HyperSpy.[60]

**Acknowledgements**

We thank the Engineering and Physical Sciences Research Council for support. We thank T. Hopper for proofreading. J. Z. thanks the China Scholarship Council for a PhD scholarship (No. 201503170255). Q.G. is grateful for financial support from the China Scholarship Council and Cambridge Trust. B.K., X.J.W. and Y.S.C. gratefully acknowledge the financial support from NSFC (91633301) and MoST (2014CB643502) of China. The work of M.H.F. and B.E. is part of the Netherlands Organization for Scientific Research (NWO). F.U.K. acknowledges scholarship funding from the Jardine Foundation. C.D. and G.D. acknowledge funding from ERC under grant number 25961976 PHOTO EM and support from the EU under grant number 77 312483 ESTEEM2. A.A.B. and D.C. are Royal Society University Research Fellows (UF130278). Y.V. gratefully acknowledges funding from the European Research Council (ERC) under the European Union's Horizon 2020 research and innovation programme (ERC Grant Agreement No. 714067, ENERGYMAPS). This project has also received funding from the European Research Council (ERC) under the European Union's Horizon 2020 research and innovation programme (Grant Agreement No. 639750, VIBCONTROL).

Recently, sequential deposition of donor and acceptor layers has demonstrated to be an alternative method to fabricate highly efficient bulk-heterojunction organic solar cells. A simple "needle" model to simulate its morphology indicates a different morphological requirement which rationalizes the high exciton dissociation efficiency.

**Keyword** organic solar cells; non-fullerene accepters; exciton dissociation;

*Jiangbin Zhang, Moritz H. Futscher, Vincent Lami, Felix U. Kosasih, Changsoon Cho, Qinying Gu, Aditya Sadhanala, Andrew J. Pearson, Bin Kan, Giorgio Divitini, Xiangjian Wan, Daniel Credgington, Neil C. Greenham, Yongsheng Chen, Caterina Ducati, Bruno Ehrler, Yana Vaynzof, Richard H. Friend, and Artem A. Bakulin\**

**Title** Sequentially Deposited versus Conventional Nonfullerene Organic Solar Cells: Interfacial Trap States, Vertical Stratification, and Exciton Dissociation

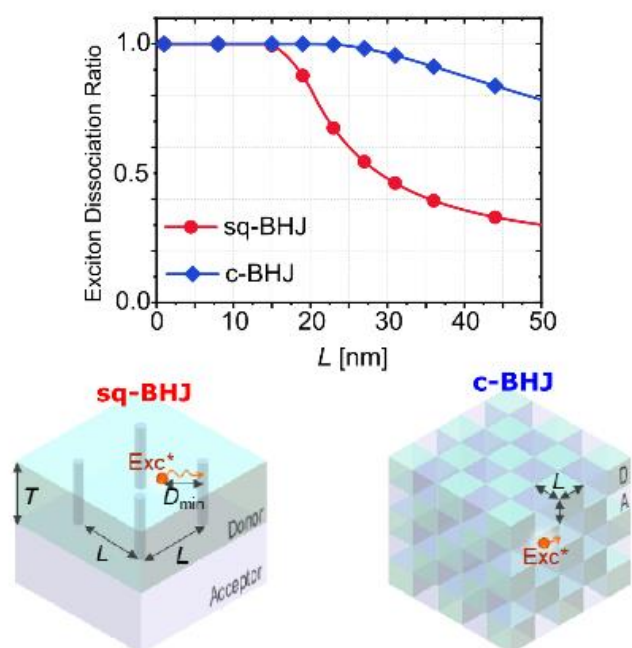